%% file: ML.tex
\documentclass{aastex63}
\usepackage{xcolor}

\received{July 27, 2021}
\accepted{November 22, 2021}
\submitjournal{\pasp}

\graphicspath{{./}{figures/}}

\begin{document}

\title{Classification of Astrophysics Journal Articles with Machine Learning to Identify Data for NED}

\author[0000-0001-9152-6224]{Tracy X. Chen}
\affiliation{Caltech/IPAC-NED, Mail Code 100-22,
Caltech,
1200 E. California Blvd., 
Pasadena, CA 91125, USA}

\author[0000-0002-9500-8587]{Rick Ebert}
\affiliation{Caltech/IPAC-NED, Mail Code 100-22,
Caltech,
1200 E. California Blvd., 
Pasadena, CA 91125, USA}

\author[0000-0002-8204-8619]{Joseph M. Mazzarella}
\affiliation{Caltech/IPAC-NED, Mail Code 100-22,
Caltech,
1200 E. California Blvd., 
Pasadena, CA 91125, USA}

\author{Cren Frayer}
\affiliation{Caltech/IPAC-NED, Mail Code 100-22,
Caltech,
1200 E. California Blvd., 
Pasadena, CA 91125, USA}

\author{Scott Terek}
\affiliation{Caltech/IPAC-NED, Mail Code 100-22,
Caltech,
1200 E. California Blvd., 
Pasadena, CA 91125, USA}

\author{Ben H. P. Chan}
\affiliation{Caltech/IPAC-NED, Mail Code 100-22,
Caltech,
1200 E. California Blvd., 
Pasadena, CA 91125, USA}

\author[0000-0002-6877-7655]{David Cook}
\affiliation{Caltech/IPAC-NED, Mail Code 100-22,
Caltech,
1200 E. California Blvd., 
Pasadena, CA 91125, USA}

\author{Tak Lo}
\affiliation{Caltech/IPAC-NED, Mail Code 100-22,
Caltech,
1200 E. California Blvd., 
Pasadena, CA 91125, USA}

\author[0000-0002-2055-7549]{Marion Schmitz}
\affiliation{Caltech/IPAC-NED, Mail Code 100-22,
Caltech,
1200 E. California Blvd., 
Pasadena, CA 91125, USA}

\author[0000-0002-4788-9236]{Xiuqin Wu}
\affiliation{Caltech/IPAC-NED, Mail Code 100-22,
Caltech,
1200 E. California Blvd., 
Pasadena, CA 91125, USA}

\begin{abstract}

The NASA/IPAC Extragalactic Database (NED) is a comprehensive online service that combines fundamental multi-wavelength information for known objects beyond the Milky Way and provides value-added, derived quantities and tools to search and access the data. The contents and relationships between measurements in the database are continuously augmented and revised to stay current with astrophysics literature and new sky surveys. The conventional process of distilling and extracting data from the literature involves human experts to review the journal articles and determine if an article is of extragalactic nature, and if so, what types of data it contains. This is both labor intensive and unsustainable, especially given the ever-increasing number of publications each year. We present here a machine learning (ML) approach developed and integrated into the NED production pipeline to help automate the classification of journal article topics and their data content for inclusion into NED. We show that this ML application can successfully reproduce the classifications of a human expert to an accuracy of over 90\% in a fraction of the time it takes a human, allowing us to focus human expertise on tasks that are more difficult to automate.

\end{abstract}

\keywords{astronomical databases: miscellaneous --- methods: analytical}

\section{Introduction} \label{sec:intro}
The NASA/IPAC Extragalactic Database (NED\footnote{\url{https://ned.ipac.caltech.edu/}}) is a comprehensive database of multi-wavelength data for extragalactic objects that provides a systematic, ongoing fusion of information integrated from hundreds of large sky surveys and tens of thousands of research publications. 
The database content includes both published information for distinct astrophysical objects such as nomenclature, positions, redshifts, photometry, diameters, images and spectra, and derived quantities such as estimates of Galactic extinction, reference-frame velocity corrections to redshifts, sizes and luminosities \citep{1988ESOC...28..335H, 2020ASPC..527....3M}.

In order to better support scientists, educators, and observatories in planning new observations, performing data analysis and modeling, making new discoveries, and preparing publications of new results, it is essential to keep the database content as current as possible. To achieve this, NED routinely reviews, extracts, and integrates data from high impact journals including Astronomy \& Astrophysics (\aap\footnote{\url{https://www.aanda.org/}}), the Astronomical Journal (\aj\footnote{\url{https://iopscience.iop.org/journal/1538-3881}}), the Astrophysical Journal (\apj\footnote{\url{https://iopscience.iop.org/journal/0004-637X}}), the Astrophysical Journal Letters (\apjl\footnote{\url{https://iopscience.iop.org/journal/2041-8205}}), the Astrophysical Journal Supplement (\apjs\footnote{\url{https://iopscience.iop.org/journal/0067-0049}}), the Monthly Notices of the Royal Astronomical Society (\mnras\footnote{\url{https://academic.oup.com/mnras}}) and Letters\footnote{\url{https://academic.oup.com/mnrasl}}, and Nature\footnote{\url{https://www.nature.com/}}. Until recently, a human expert would first scan the articles and determine if the topic of an article is appropriate to be included in the database, i.e., if the article contains data for extragalactic objects. When the topic is considered relevant, the expert would further review the content of the article to determine which of the different types of data should be extracted and integrated into the database. This is a very labor-intensive process. In 2018 alone, NED reviewed a total of 14,123 journal articles, which translated to an average of $\sim$270 articles per week. Given the increasing volume of publications and the growing volume and complexity of data therein, it is neither efficient nor sustainable to rely only on human effort to continue this work. We therefore explored applying machine learning (ML) technique to classify literature with the goals of automating and accelerating the process of selecting journal articles and data for inclusion into NED while maintaining a high level of consistency with human classifications.

This paper is organized as follows. \S\ref{sec:method} describes the method and software we selected for our ML application, and how we implemented it in our use case. \S\ref{sec:topic} discusses the sample curation and training for selecting journal articles that are of interest to NED, the performance of our best model validated against the testing sets, and its deployment into the NED production pipeline. In \S\ref{sec:content}, we extend the approach to further identify the different data content in the articles, e.g., astrometry, photometry and redshift data. We summarize our study and the results in \S\ref{sec:conclusion}.

\section{Method} \label{sec:method}
\subsection{Algorithm}\label{subsec:algorithm}
ML algorithms are generally divided into two groups: supervised and unsupervised. The classification of NED relevant articles and data calls for a supervised learning algorithm as we want to learn the mapping of features in journal articles to the decisions of human experts over the last 10-20 years, and use the learned relationship to predict the decisions on new, previously-unseen articles. We expected to have a high dimensionality in feature space ($\sim$ 1M features) as the input consists of fairly large documents ($\sim$ 0.1-1M bytes per article), and a fairly large number of them ($\sim$ several thousand articles). 
Hence, we chose to focus on leveraging existing open-source software that is well-reputed to build models over text data. Additionally, the ML solution and feature extractor need to operate in a pipeline/background mode without graphical interfaces, remote connections, or human interaction. 
After careful study, we chose the Stanford Classifier (SC\footnote{\url{https://nlp.stanford.edu/software/classifier.shtml}}) developed by the Stanford Natural Language Processing Group. The SC implements a Maximum Entropy classifier with Conjugate-Gradient Descent optimization (MaxECG), which is described in detail in \citet{SC2003}. The program includes an easy-to-use command-line interface with several tunable algorithms, as well as a feature extraction and mapping/encoding component. The software is well supported with a strong user base, has an open source licensing under the GNU General Public License\footnote{\url{http://www.gnu.org/licenses/old-licenses/gpl-2.0.html}}, and has also been updated and improved over the years. For the work described here, we used version 3.9.2, which was released in 2018. It is worth pointing out that our goal was not to perform a computer science experiment in search of a ``best" solution, but to apply a well-tested package that has been successfully used in other application domains. For example, the Max Entropy (MaxEnt) SC is one of four leading supervised text classification algorithms that were compared for sentiment analysis of movie recommendations in Twitter feeds \citep{Armentano2015}.

\subsection{Overview of ML Processing for NED}\label{subsec:process}
\begin{figure}[ht!]
\plotone{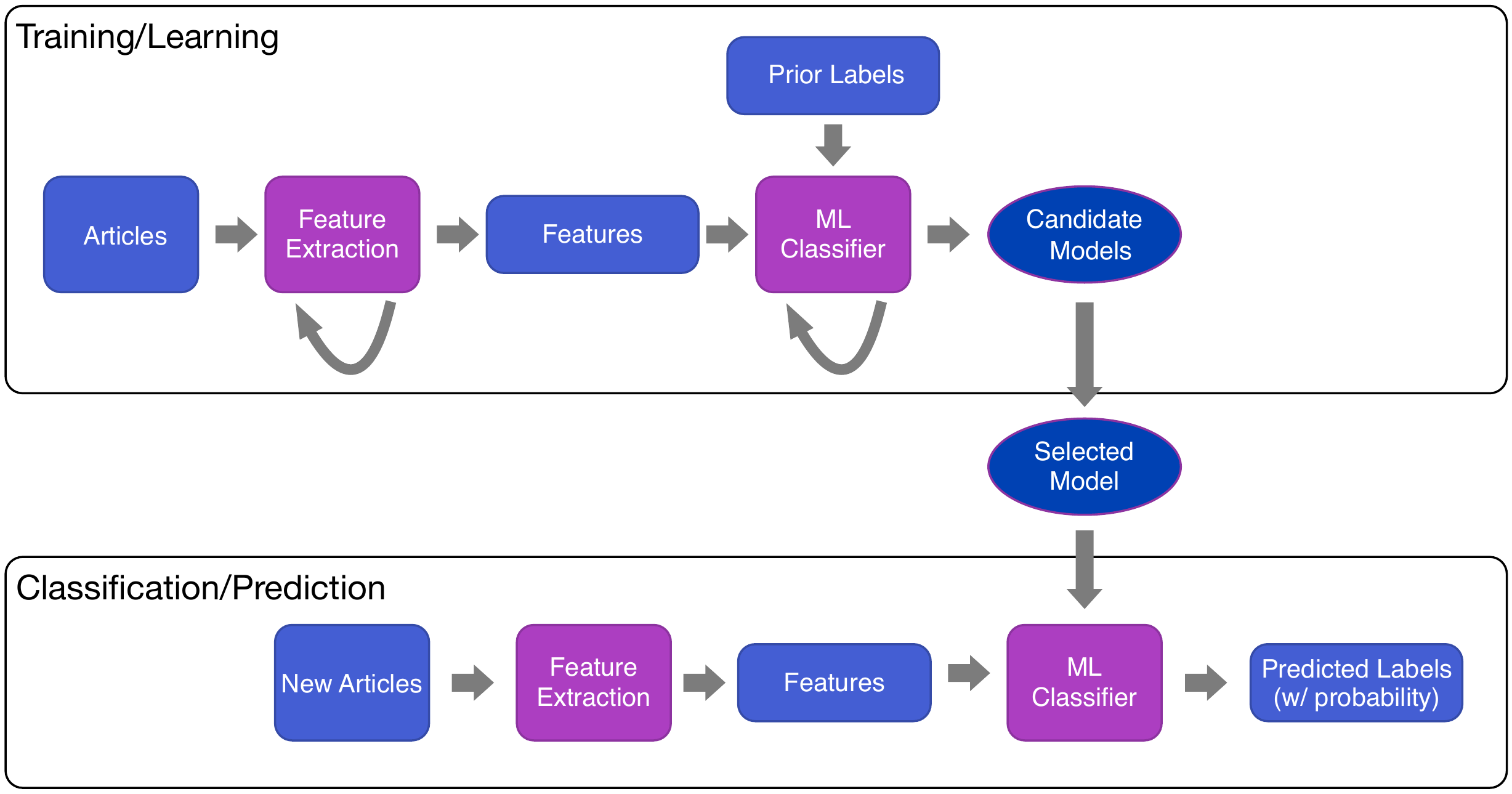}
\caption{Machine learning journal article classification process in NED.\label{fig:ml}}
\end{figure}

The ML process involves two major cycles: the learning (or training) cycle and the prediction cycle. Figure \ref{fig:ml} illustrates the general article classification process in NED.

In the learning cycle, models are constructed and evaluated using training sets and test sets that are independent subsets of the input data. Our input data are the HTML files of articles published in the astronomical journals (see \S \ref{sec:intro}), acquired from the journal websites via their ADS links, and the labels placed by human experts which we adopt as ground truth. The Stanford Classifier is used to perform feature characterization, identification and extraction on the training sets. Features can be words, parts of words, stems of words
, characters, groups of characters (letter-n-grams\footnote{An n-gram is any ordered collection of n symbols, see \url{https://en.wikipedia.org/wiki/N-gram}. Letter-n-grams use letters as the unit. Similarly, word-n-grams use words as the unit.}), and groups of words (word-n-grams). We experimented with different feature models and found the ``Chris4” word-shape\footnote{Word-shape is the pattern of characters and delimiters that are considered ``a word" by the feature extractor. Several of these, like Chris4, are built-in to the SC software, but others can be created using common regular-expression syntax. See \url{https://nlp.stanford.edu/nlp/javadoc/javanlp/edu/stanford/nlp/classify/ColumnDataClassifier.html} for more details.}, with leading or trailing 2-5 letter-n-grams, as well as whole-words, performed the best. We also learned that feature extraction is the most expensive computational activity, about 10 times the demand of the ML training itself. We therefore extract features from each article in a training set, combine them into an intermediate catalog of feature sets, and reuse them in subsequent training exercises. Positive or negative weights of the features for each class are learned in the training process based on the priors. Table \ref{tab:feature} presents some examples of the features with the weights learned in the training process. It is noticeable that the weights of the features are sensitive to the letter cases. In the examples here, the words ``galaxy”, ``Galaxy”, ``GALAXY” and ``GALaxy" are shown to have very different weights.

\input{tables/feature}

Multiple candidate models are trained and examined by varying the choice of the training and test sets, from which we select the best-performing model. The primary evaluation metrics we use in accessing the performance of the models are the classification accuracy and recall. Accuracy is defined as the fraction of the correct predictions (i.e., consistent with the decisions by the human experts) to the total number of predictions made: 

\begin{equation}
Accuracy = \frac{T_p + T_n}{T_p + T_n + F_p + F_n} .
\end{equation}
Here $T_p$ represents the true positives where the model correctly predicts the positive class, $T_n$ represents the true negatives where the model correctly predicts the negative class, $F_p$ represents the false positives where the model incorrectly predicts the positive class, and $F_n$ represents the false negatives where the model incorrectly predicts the negative class.
Recall is defined as the fraction of actual positives that are identified correctly:

\begin{equation}
Recall = \frac{T_p}{T_p + F_n}.
\end{equation}
The initial requirement for our ML application was to minimize the human effort and accelerate the rate of classification while maintaining 90\% consistency with human classifications. By looking at the instances where the ML approach differ from the human decisions, we found that classifications by our human expert, though very accurate, typically have an error rate of $\sim$ 1\%. This led us to set our goal as maximizing the efficiency of classification at a high accuracy rate ($\sim$ 90\%) and an acceptable recall rate ($\sim$ 80-90\%).  

In the prediction cycle, we extract features from new, previously unseen articles and feed them into the selected model. The model then returns a predicted label for the new article with a probability indicating the likelihood of the prediction. 
In \S\ref{sec:topic} and \S\ref{sec:content}, respectively, we describe specific applications of this methodology to successfully identify articles appropriate for NED and to classify the types of data contained withing the papers (object positions, photometry, redshifts, etc.), helping expedite the data ingestion from the literature into the database.

We note that, because the classifier is trained on past publications and used to predict future publications (domain shift), it is better to accept a slightly higher rate of false positives because all the positives will be further inspected by human experts in order to clean, reformat, and validate the data for ingestion into the database. During that process, the false positives will be effectively removed and not contaminate the end result. This still saves substantial time that used to be spent examining the true negatives, allowing human resources to work on the more difficult-to-automate data validation and extraction tasks.
We are aware that NED will inevitably suffer from some incompleteness due to the false negatives. However, the main goal for this ML approach is to maximize efficiency while keeping as current as possible with the increasing flow of data from the literature. Some omissions are acceptable with an appreciable gain of processing more data into NED more quickly overall, which will be a greater benefit to a larger number of users. In addition, authors/users are generally quick to inform us if there is an important omission, and the missed data can always be added afterwards.

\section{NED-Appropriate Topics Classification}\label{sec:topic}
Our first implementation of this ML approach is to identify if the topic of an article is relevant to NED or not, i.e., NED-Appropriate Papers (NAPs) or (Not NED-Appropriate papers (NNAs). Historic data show that among the journals we reviewed, about a quarter to a third of the articles published each year contain extragalactic objects and therefore need to be selected and further categorized for inclusion into the NED database.    

\subsection{Training and testing}
We selected training and testing sets from 16 years (2003-2018) of AJ articles. There were 6,987 articles in total, with 1,984 NAPs and 5,003 NNAs based on human classifications. We acquired the HTML files of these articles through the ADS API\footnote{\url{https://ui.adsabs.harvard.edu/help/api/}}. We then extracted the features from the raw HTMLs, i.e., no stripping of the HTML formatting, as this saves considerable time in processing.
For each article we extracted on the order of a million features, and it took a few days of CPU time in total to generate all the feature sets for the 6,987 articles. 

From the 16 years of AJ articles, we created 100 training/testing iteration samples, 
each containing the 1,984 NAPs and 1,984 randomly selected articles from the 5,003 NNAs. 
Equal number of positives and negatives were chosen to avoid an imbalanced data set, which may result in better performance on the larger class and worse performance on the smaller class, and may also affect the use of regular accuracy as a evaluation metric \citep{2019arXiv190407248B}. For each iteration, we processed 10 folds of the SC-MaxECG algorithm on a randomly selected 90\% of the articles for training and 10\% for testing. Essentially, this created and tested 1,000 models with different choices of training and test sets. The resulting classifier models typically selected 720,000 features of significance and scored an accuracy ranging from 89\% to 97\%, and a recall ranging from 92\% to 96\% when evaluated on the test sets. The 100 rounds with 10 folds of ML computations used 22 CPU hours in total.

\begin{figure}[ht!]
\plotone{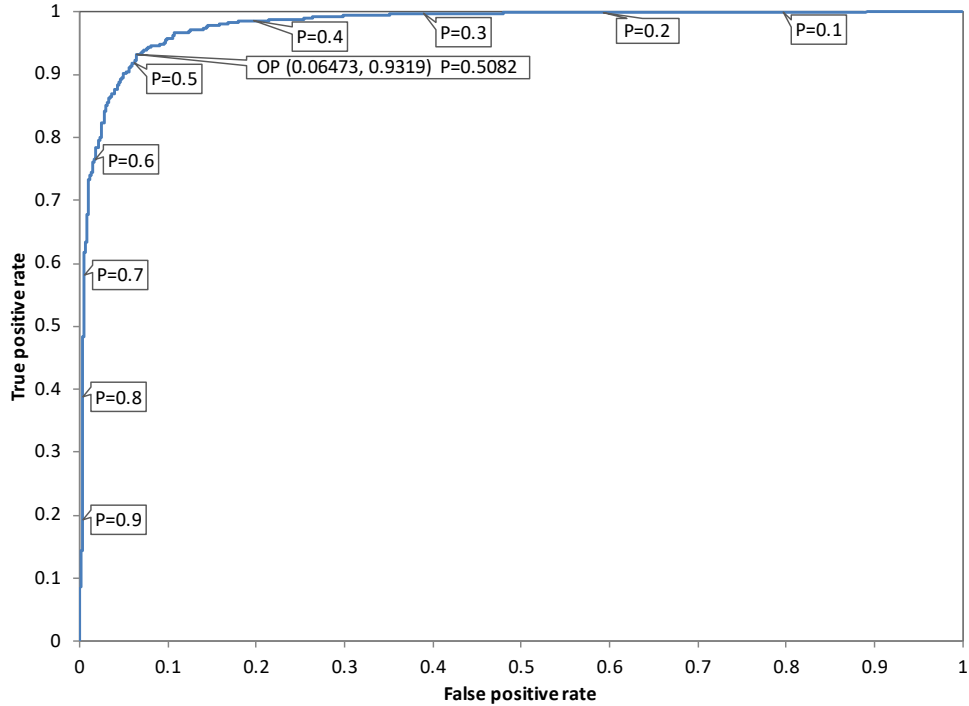}
\caption{Receiver Operating Characteristic (ROC) curve for the NAP/NNA classifier implemented in NED. Different probability points are labelled as well as the optimum point (OP) where the classifier discriminates the maximum true positive rate for the least false positive rate. Here the OP occurs at a true positive rate of 0.9319 and a false positive rate of 0.06473, which corresponds to a probability threshold of p = 0.5082, very close to the default p = 0.5 that we adopted.}
\label{fig:roc}
\end{figure}

Table \ref{tab:output} provides an example output generated by one of the ML models. For each article (represented by its ADS Bibcode), the classifier outputs a label for the article, and a probability of this article belonging to this class. For example, the classifier decided that article 2012A\&A...537A..29R is 90.8\% likely to be a NAP and article 2017ApJ...841L..16V is 99.8\% likely to be a NNA. 
We adopted the default probability threshold of 0.5 in SC as our decision cutoff, i.e., an article would be labelled NAP if p(NAP) $>$ 0.5 and NNA if p(NNA) $>$ 0.5.
We found that this threshold of p = 0.5 corresponds closely to the optimum point where the difference between the true positive rate and the false positive rate is maximized for the NAP/NNA classifier. 
This is illustrated in Figure \ref{fig:roc} using the Receiver Operating Characteristic (ROC) curve \citep{2006PaReL..27..861F, Tharwat2021} of the best trained model, where the true positive rate is plotted against the false positive rate and the optimum point is shown to occur at a threshold of p = 0.5082. Additionally, the predictions made with this cutoff threshold do meet our goal for accuracy and recall.

\input{tables/output}

During the testing, we found that 455 articles were classified opposite to NED’s historical choice (i.e., false positives and false negatives) in at least one out of the 100 iterations, and 137 articles classified opposite to NED’s choice in all 100 iterations. This can mean there are mistakes in the prior classification, but as the ML is ``learning" from the priors, it more likely indicates that there is an inconsistency (in the vocabulary, syntax, or both), or simply not enough information (distinguishing terms/features).

When applying the best trained model on the entire corpus of AJ articles, we were able to reproduce the human classifications to an accuracy of 97\% and a recall of 94\%. We further validated the model on other journals from the same period of time (2003-2018), and were able to reproduce the priors to an accuracy of 93\% for ApJ and ApJL, 91\% for ApJS, and 90\% for MNRAS and MNRAS Letters, individually, with a recall of $>$ 91\%  for all these journals. 

We also experimented, but not elaborately, with the alternative Naive Bayes (NBy) classifier available in the SC package. NBy classifiers are known to compete favorably with MaxEnt classifiers given large training sets, i.e., $\gg$ 10,000 items in the corpus; MaxEnt classifiers perform best at general text classification with small corpora, i.e., $\lesssim$ 10,000 \citep{Potts2011, Yuret2010, 2002cs........5070P}. We evaluated the performance using computational speed, accuracy and recall. We found that MaxEnt does become slightly more expensive in computation as the number of training items increases. However, at the scale of the NED literature set (1000's), this is easily overcome by computing facilities at hand. The NBy method is problematic for small training sets, as accuracy and recall fall off quickly. In the few experiments conducted, the accuracy of the NBy classifier ranged from 40\% to 70\%, with no noticeable improvement in speed. 

Other areas that were examined include: (1) Using a simpler word-shape, and/or not using letter-n-grams. This reduces the number of features substantially (factor of $\sim$ 2/3) and therefore computation. However, accuracy and recall suffered immensely, falling by more than 30\%; (2) Using only defined keywords, known by experience to be related to the selection. We confirmed that the classifiers generally did select features that include forms of the NED keywords historically used by humans in skimming articles for significant information. It was in this process where we found that letter-case was significant in weighting such features.  

\subsection{Deployment to production} \label{subsec:ncsa1}
The best performing model, dubbed SC-classifier-AJ-PF1984-064 (trained on 1984 positive priors and like number of negative priors of AJ articles), was selected for deployment in the NED production pipeline and has been used to automate the identification of NED-appropriate papers among new articles published in the astronomical journals since early 2020. 

\begin{figure}[ht!]
\epsscale{0.8}
\plotone{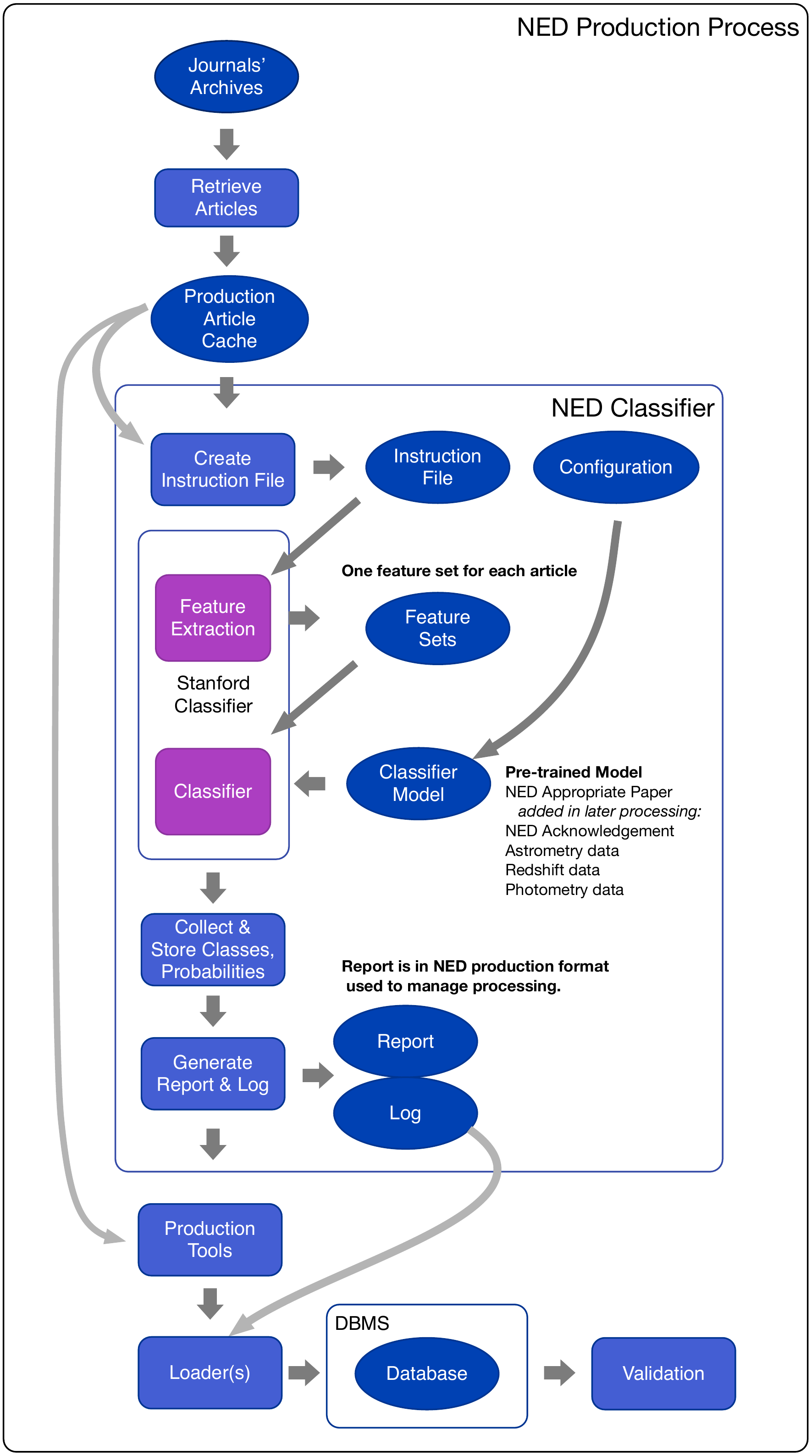}
\caption{The acquisition, evaluation and loading process in the NED production pipeline, with the ML classifier highlighted. 
\label{fig:prod}}
\end{figure}

Figure \ref{fig:prod} illustrates the acquisition, evaluation, and loading process in the NED production, with the ML classifier highlighted. The tool takes a list of files containing the pre-downloaded article texts (in HTML obtained from publishers' on-line archives), constructs the input needed to inform the feature extractor, invokes the feature extractor and the classifier with the configured model on the extracted features, and outputs the report and log files. The former records only the positive classifications that will be used as input to NED's existing loader program. The latter records all the classifications and the computed likelihood values, which are reserved for future testing and tuning. 

Since the performance of the selected model may degrade due to the changing of input data (e.g., shifts in research topics or journal formats can introduce new features), periodic model retraining is necessary to stay current with the astronomical literature. We set a periodic training schedule of once per year. The new training sets are from human classifications on one or two volumes per journal every 6 months, and the collected false positives detected in the data preparation stage. We deployed our model into NED production in early 2020 and collected our first batch of new human classifications in the summer of 2020. We compared the labels assigned by human experts with the predictions from the ML classifier, and present our results in Table \ref{tab:ncsa1}. It is shown that although the model was trained on pre-2019 AJ articles, the performance is still strong: the overall accuracy and recall are both at 92\% for articles published in 2020. 
We plan to incorporate these new training sets in 2021/2022. Our short-term focus, however, is to expand the scope of the ML application to cover classifying types of data in the NAP articles.

\input{tables/ncsa1}


\section{Data Content Classification} \label{sec:content}
The success of the ML application to automate and accelerate the selection of NED relevant articles from astronomical journals encouraged us to further extend the approach to include the data content classifications, i.e., what types of data an article has to offer. The general methodology was the same as in our previous implementation, but was now applied to NED ``data-frames” for astrometry (positions), kinematics (redshift), photometry, and in-literature acknowledgement of NED, in addition to the appropriateness for an article to be included in NED (NAPs).

The types of data NED routinely integrates into the database contain more than the selected few above. These categories were chosen mostly due to the fact that they have the largest sample size in the database. Additionally, positions, photometry and redshift are among the most fundamental properties NED keeps for an object, and acknowledgement is an important metric used as a basis for productivity and impact by data archives such as NED. 

\subsection{Training and testing}
We initially assembled a sample containing articles from all the NED core journals (A\&A, AJ, ApJ, ApJL, ApJS, MNRAS and MNRAS Letters) between the year 2004 and 2019. The NED database management system (DBMS) was used to randomly select training and test sets, based on desired criteria (NAPs or NNAs; contains specific data types or not). However, during the training/testing process, we found that articles published before 2011 were more prone to classification error. Upon examination, we found that two issues were involved. One was a significant format change in some journals’ publication around 2010 (e.g., A\&A and MNRAS), which created some variations in features that compromised the performance of the model. The other was due to the fact that there were less data in NED for the years before 2011, compared to the years after. Since our goal is to have the model predict future instances accurately, it is therefore important to make sure the data the model is trained on are similar to what is expected in the near future. Hence, we revised our sample selection to include only the articles from 2011 through 2019. Table \ref{tab:sample} shows the total size of the final sample, together with the number of positive priors in each of the data categories to be modelled. It is worth pointing out that in the sample selection the NAP and acknowledgement classifications are independent, i.e., an article can acknowledge the use of NED without providing any data on objects, hence is not a NAP. The classifiers on different data types (astrometry, photometry, and kinematics) are only trained on samples that are NAP-condition-positive; that is, for example, the astrometry training set (both positives and negatives) is taken from the NAP subset. 

\input{tables/sample}

For each of the 5 categories (NAP, acknowledgement, astrometry, photometry, and kinematics), we processed 20 iterations of the SC-MaxECG on randomly selected subsets from the entire sample. Each subset contains 90-95\% of the positives and an equal number of negatives in each category for training, and 5-10\% of the positives and negatives for testing. All together, we trained 100 classifier models, and it took 144 CPU hours (not including feature extraction time). Since feature extraction of individual articles and the concatenation of different feature sets from all the articles in the sample are very time-consuming, on the order of days in CPU time, we chose to do it once at the beginning and used the same combined feature file in the training of all five classifiers. 

The resulting models from the 20 iterations have an accuracy range of 91-93\%, 96-98\%, 87-93\%, 88-92\% and 88-92\% for NAP, acknowledgement, astrometry, kinematics and photometry classifiers, separately. Table \ref{tab:model} presents, for the best-performed model of each classifier, the total number of articles and the number of positive priors in the subset, together with the accuracy and recall when evaluated against the respective testing set. The accuracy rates are at $>$92\% for all the classifiers, and the recall rates are at $>$93\% except for the acknowledgement category. In the case of acknowledgements, there was a shortage of positive priors due to a shift in the definition of positives for this category. ``Acknowledgement of NED” was initially defined as any mention of NED in the article text using ADS keyword extraction, or classification by ADS as having cited NED, or classification by NED team through visual inspection. The training sample was selected using the initial definition. However, we later revised the definition to only consider the NED team confirmed acknowledgements or citation, which resulted in a smaller training set and subsequently a somewhat limited yet still sufficient accuracy and recall in testing. 

\input{tables/bestmodel}

\subsection{Deployment to production} \label{subsec:ncsa3}
We integrated the above-mentioned best-performing models to production at the beginning of 2021. Our initial approach was to remove the articles identified as NNA from the input of the data content classifiers, so that the data type classification would only occur on articles that were already found to be NAP. However, we found that it was considerably faster to run the entire input set (including both the NAPs and NNAs) through all classifiers, and generate recommendations/classifications based on the combined output. We therefore augmented the program to run all five classifiers simultaneously on each article, extract results and output them in the required format of the NED loader program. 
During the initial deployment, we randomly selected a few volumes from different journals and had our human experts label the articles in order to assess the performance of the models. Table \ref{tab:2021} provides a few examples of the results of classification accuracy based on post-classification human verification. The total number of articles as well as the number of NAPs in each volume are listed here to caution that these are all small samples, hence the accuracy and recall rates are very sensitive to the mis-classification of even a few articles. For example, ApJ 914 contains only four articles with NED acknowledgement in the entire volume. The ML approach mis-classified three, resulting in a low recall of 25\%. Nonetheless, the classifiers are shown to perform well outside of their training domain (year 2011-2019) on these articles published in 2021. As of September 2021, this ML classifier has been in full operations for 8 months, and has reduced $\sim$ 80\% of the labor for the content classification task. This translated to a team gain of 0.4 Full-Time Equivalent (FTE) which was then focused on tasks that cannot be automated.

\input{tables/ncsa3}

As has been discussed previously in \S\ref{subsec:ncsa1}, these classifiers are trained on past publications and are being applied to publications that are new, with information/features that sometimes have not been seen before. Therefore, it is important to monitor the trend of accuracy and recall and be prepared to refresh the models with more current publications. Currently, we are keeping the periodic retraining schedule of once per year. The schedule may be changed based on the trend of the performance, which we are assessing every 6 months using human classified labels on randomly selected journal articles.

\section{Conclusion}\label{sec:conclusion}
We explored the use of ML techniques to process articles in the astronomical literature, selecting those that are appropriate for inclusion into NED, and categorizing the different types of data contained in the articles for further extraction and ingestion. We showed that this automated process can reproduce the historic classifications by human experts of articles relevant to NED and selected data types (astrometry, photometry, kinematics and acknowledgement) to an accuracy of 92\% or greater. It is worth noting that this ML approach does not completely remove human involvement in the process. Human expertise and learning are needed for marginal cases that are not resolved by existing capabilities. Further, the trained model is applied to a domain-shifted data set for predictions. Newly-emerged data content, variations in formats and representations can all contribute to deteriorating performance over time. Hence, there needs to be a continuing education program for retraining and updating the classifier models with new literature, which will require new labels identified by human experts periodically. 

Overall, we have demonstrated the feasibility and value of using ML to reduce the human effort in this historically labor-intensive task while maintaining a high level of accuracy and consistency. This ML application on both subject and content classifications has been in operations since January 2021, and has gained the team about 0.4 FTE, i.e., the equivalent of 4 person-years of labor over a decade of NED operations, that we have refocused on tasks that have little or no hope of being handled with artificial intelligence or machine learning anytime soon, such as analyzing and understanding the data content well enough to clean, reformat, apply or validate metadata to prepare input to the database.  


\acknowledgments
We want to acknowledge the past and continuing work of the Natural Language Processing Group at Stanford University, without which this application would have been much more difficult. We are also grateful to the American Astronomical Society Journals and IOP Publishing, the Oxford University Press, and EDP Sciences for their support in providing extensive and on-going access to their publications. We thank the anonymous referee for the critical review and comments which resulted in substantial improvements to the article. This work was funded by the National Aeronautics and Space Administration through a cooperative agreement with the California Institute of Technology.

\facilities{ADS, NED}

\software{Stanford Classifier v3.9.2, The Stanford Natural Language Processing Group, \url{https://nlp.stanford.edu/software/classifier.shtml} 
}

\bibliography{ML}{}
\bibliographystyle{aasjournal}

\end{document}

%% file: tables/feature.tex
\begin{deluxetable*}{lcl}
\tablecaption{Example features and weights extracted from a training sample\label{tab:feature}}
\tablewidth{0pt}
\tablehead{
\colhead{Feature} & \colhead{Weight} & \colhead{Note}
}
\startdata
galaxy & 0.2875 & Whole word \\
galax & 0.3084 & Leading 5 letter-n-grams \\
xies & 0.3237 & Trailing 4 letter-n-grams \\
GALAXY & 0.0957 & Whole word \\
Galaxy & 0.0370 & Whole word \\
GALaxy & 0.0001 & Whole word \\
\enddata
\end{deluxetable*}

%% file: tables/output.tex
\begin{deluxetable*}{lcc}
\tablecaption{Example output from a NAP/NNA classifier model}\label{tab:output}
\tablewidth{0pt}
\tablehead{
\colhead{Bibcode} & \colhead{Label} & \colhead{Probability\tablenotemark{a}}
}
\startdata
2012A\&A...537A..29R & NAP & 0.908 \\
2014MNRAS.444.1862S	& NAP & 0.992	\\
2018MNRAS.475.5513V	& NAP & 1.000\\
2018ApJ...868L..15S	& NNA & 0.998	\\
2015ApJ...799..237U	& NNA & 0.882	\\
2016ApJ...821L..12D	& NAP & 1.000	\\
2013ApJ...763...19R	& NAP & 1.000	\\
2016MNRAS.459.3432M	& NNA & 0.559	\\
2018MNRAS.473.5237K	& NNA & 0.702	\\
2011MNRAS.411.2426G	& NNA & 0.999	\\
2017ApJ...841L..16V	& NNA & 0.998	\\
\enddata
\tablenotetext{a}{This is the likelihood of an article to be classified as one type or the other. We require a probability value to be $> 0.5$ in order to label an article NED-Appropriate Paper (NAP) or Not NED-Appropriate paper (NNA).}
\end{deluxetable*}

%% file: tables/ncsa1.tex
\begin{deluxetable*}{lccccccc}
\tablecaption{Performance of the topic classifier on 2020 articles\label{tab:ncsa1}}
\tablewidth{0pt}
\tablehead{
\colhead{} & \colhead{ApJS 249} & \colhead{A\&A 638} & \colhead{A\&A 639} & \colhead{MNRAS 494} & \colhead{AJ 160} & \colhead{ApJ 895} & \colhead{Overall} \\
\colhead{} & \colhead{} & \colhead{} & \colhead{} & \colhead{} & \colhead{Issue 1, 2} & \colhead{} & \colhead{}
}
\startdata
Number of articles & 36 & 178 & 160 & 476 & 96 & 200 & 1,146 \\
Accuracy & 97\% & 94\% & 96\% & 89\% & 100\% & 87\% & 92\% \\
Recall & 89\% & 100\% & 98\% & 96\% & 100\% & 71\% & 92\% \\
\enddata
\end{deluxetable*}

%% file: tables/sample.tex
\begin{deluxetable*}{lr}
\tablecaption{The Sample\label{tab:sample}}
\tablewidth{0pt}
\tablehead{
\colhead{Category} & \colhead{Number of articles}
}
\startdata
Total & 83,755 \\
NAP & 21,820 \\
Astrometry & 6,660 \\
Photometry & 9,040 \\
Kinematics & 7,248 \\
Acknowledgement & 5,501 \\
\enddata
\tablecomments{This table gives the total number of articles in A\&A, AJ, ApJ, ApJL, ApJS, MNRAS and MNRAS Letters from 2011 to 2019, and the number of positive priors in each of the data categories.}
\end{deluxetable*}

%% file: tables/bestmodel.tex
\begin{deluxetable*}{lrrrrr}
\tablecaption{Input and performance of the best classifier in each data type category\label{tab:model}}
\tablewidth{0pt}
\tablehead{
\colhead{Classifier} & \colhead{NAP} & \colhead{Acknowledgement} & \colhead{Astrometry} & \colhead{Redshift} & \colhead{Photometry}
}
\startdata
Sample Size\tablenotemark{i} & 41,458 & 10,450 & 13,654 & 13,756 & 17,176 \\
Positives & 20,704 & 886 & 6,317 & 6,891 & 8,578 \\
Accuracy & 93\% & 98\% & 93\% & 92\% & 92\% \\
Recall & 93\% & 80\% & 96\% & 93\% & 94\% \\
\enddata
\tablenotetext{i}{This is the size of the subset used to train the classifier, containing 90-95\% of the positive priors and an equal number of negative priors randomly selected from the total sample in each category for training.}
\end{deluxetable*}

%% file: tables/ncsa3.tex
\begin{deluxetable*}{llccccc}
\tablecaption{Performance of the new data-type classifier on some 2020 and 2021 journal articles.\label{tab:2021}}
\tablewidth{0pt}
\tablehead{
\colhead{Journal} & \colhead{} & \colhead{NAP} & \colhead{Acknowledgement} & \colhead{Astrometry} & \colhead{Redshift} & \colhead{Photometry}
}
\startdata
A\&A 645 & Accuracy & 96\% & 99\% & 91\% & 93\% & 91\% \\
(160:38) & Recall   & 89\% & 88\% & 77\% & 91\% & 82\% \\
\\
A\&A 651 & Accuracy & 93\% & 96\% & 97\% & 94\% & 93\% \\ 
(138:36) & Recall   & 83\% & 62\% & 87\% & 67\% & 77\% \\
\\
ApJ 904  & Accuracy & 96\% & 99\% & 94\% & 94\% & 91\%  \\
(233:54) & Recall   & 91\% & 75\% & 100\%& 92\% & 83\% \\
\\
ApJ 905  & Accuracy & 96\% & 97\% & 93\% & 93\% & 93\% \\
(215:60) & Recall   & 92\% & 60\% & 96\% & 96\% & 95\% \\
\\
ApJ 906  & Accuracy & 95\% & 99\% & 95\% & 92\% & 90\% \\
(151:49) & Recall   & 90\% & 80\% & 100\%& 100\%& 94\% \\
\\  
ApJ 914  & Accuracy & 97\% & 98\% & 95\% & 93\% & 94\% \\
(185:59) & Recall   & 92\% & 25\% & 94\% & 88\% &100\% \\
\\
ApJS 252 & Accuracy & 97\% & 97\% & 94\% & 97\% & 81\% \\ 
(32:11)  & Recall   & 91\% & 75\% & 88\% & 89\% & 100\% \\ 
\\ 
ApJS 255 & Accuracy & 97\% & 93\% & 87\% & 93\% & 90\% \\
(30:12)  & Recall   & 100\%& 71\% & 83\% & 80\% & 100\%\\
\\
MNRAS 499 & Accuracy& 91\% & 98\% & 90\% & 93\% & 90\% \\ 
(451:151) & Recall  & 97\% & 79\% & 83\% & 87\% & 91\% \\
\enddata
\tablecomments{The total number of articles and the number of NAPs in each volume are given in the parenthesis under the journal volume name.}
\end{deluxetable*}